\documentclass[amsmath,amssymb,aps,prfluids]{revtex4}

\usepackage{graphicx}% Include figure files
\usepackage{subfig,enumerate}
\usepackage[mathscr]{euscript}
\usepackage{upgreek}
\usepackage{cancel}
\usepackage{dcolumn}% Align table columns on decimal point
\usepackage{bm}% bold math
\usepackage{hyperref}% add hypertext capabilities
\usepackage{xcolor}

\begin{document}
\newcommand{\sm}[1]{{ #1}}
\newcommand{\nd}[1]{{ #1}}
\newcommand{\ub}{\mathbf{u}}
\newcommand{\eb}{\mathbf{e}}
\newcommand{\pard}[2]{\frac{\partial #1}{\partial #2}}
\newcommand{\Pe}{\mbox{Pe}}

\title{Steady state propulsion of isotropic active colloids along a wall}
\author{Nikhil Desai}
\author{S\'ebastien Michelin}%
 \email{sebastien.michelin@ladhyx.polytechnique.fr}
\affiliation{%
 LadHyX, D\'epartement de M\'ecanique, CNRS – Ecole Polytechnique, Institut Polytechnique de Paris, 91128 Palaiseau Cedex, France\\
}%

\date{\today}% It is always \today, today,

\begin{abstract}
Active drops emit/absorb chemical solutes, whose concentration gradients cause interfacial flows driving their own transport and the propulsion of the droplet. Such non-linear coupling enables active drops to achieve directed self-propulsion despite their isotropy, if the ratio of advective-to-diffusive solute transport, i.e. the P\'eclet number $\Pe$, is larger than a finite critical threshold. In most experimental situations, active drops are non-neutrally buoyant and thus swim along rigid surfaces; yet theoretical descriptions of their non-linear motion focus almost exclusively on unbounded domains to circumvent geometric complexity. To overcome this gap in understanding, we investigate the spontaneous emergence and nonlinear saturation of propulsion of an isotropic phoretic colloid along a rigid wall, to which it is confined by a constant external force (e.g., gravity). This phoretic particle model is considered here as a limiting case for a viscous active drop. We show that, \nd{for moderate $\Pe$}, the particle motion and associated chemical transport reduce the chemically-induced wall repulsion, thereby causing the particle to swim progressively closer to the wall as $\Pe$ increases. Far from hindering self-propulsion, this reduction in the particle-wall separation is accompanied by a wall-induced efficient rearrangement of the solute concentration gradients driving the particle, thus augmenting its swimming speed.
\end{abstract}

%Use showkeys class option if keyword
\maketitle

\section{Introduction}\label{intro}

Active drops are a class of synthetic microswimmers that utilize their interfacial properties to convert chemical energy to mechanical motion~\cite{Thutupalli2011, Peddireddy2012, Herminghaus2014, Maass2016}. They display a wide variety of trajectories, e.g., chaotic~\cite{Suga2018}, curling~\cite{Kruger2016b, Suga2018}, diffusive~\cite{Izzet2020}; and swimming behaviors, e.g., mode-switching~\cite{Hokmabad2021}, chemo-sensitivity~\cite{Jin2017, jin2018}, rheotaxis~\cite{Dwivedi2021}. Due to this diversity, active drops can mimic the motion of complex biological systems, e.g., perform chemotaxis~\cite{Berg1975, Jin2017} or upstream swimming~\cite{Hill2007, Dey2021}; yet, their own motion can be explained via relatively simpler physico-chemical interactions. In addition, these drops are easy to produce~\cite{Seemann2016}, which makes them ideal for the analysis of important micro-scale hydro-chemical phenomena, like individual motion, pair interactions and large-scale collective motion.

The most important properties governing the motion of an active drop are its (i) \emph{activity}, the ability to exchange chemical solutes with its surroundings, and, (ii) \emph{mobility}, the ability to convert the solutes' non-uniform spatial distribution into interfacial flows. Unlike intrinsically asymmetric Janus particles~\cite{Ebbens2016, Moran2017}, active drops lack any built-in asymmetries and as such rely on a non-linear, symmetry-breaking instability to self propel~\citep{Izri2014}. Small disturbances in the concentration of emitted solute give rise to interfacial flows that advect the solute along the drop's surface, amplifying the initial disturbance and establishing a concentration polarity across the drop. Thus, the directional symmetry is broken and the inertialess drop swims as a response to the sustained fluid flow around it. Central to this swimming mechanism is the requirement that the advective transport of the solute must dominate its molecular diffusion, or, the characteristic P\'eclet number of the system $\Pe$, must be larger than a critical value $\Pe_c$, as confirmed in prior theoretical studies~\cite{Michelin2013, Morozov2019}. This finite advection couples the hydrodynamic and chemical fields around the drop and makes it an inherently non-linear system, from a modeling perspective.

In experiments, active drops are generally confined to rigid walls owing to a density mismatch with the surrounding fluid or restricted geometries such as Hele-Shaw cells~\cite{Moerman2017, deBlois2019, Cheon2021, Hokmabad2021}. To avoid the difficulty of solving a non-linear problem in complex geometry, theoretical analyses on the motion of isotropic active colloids have traditionally focused on unbounded flows~\cite{Yoshinaga2012, Michelin2013, Schmitt2013, Izri2014} or employed simplifications that allow linearised analysis, e.g., neglecting solute advection and only considering confinement effects~\cite{Dominguez2016, Yariv2016, Yariv2016b}. The first investigation avoiding both these assumptions considered normal/axisymmetric collisions of an active drop with a rigid wall, where it was shown that $\Pe$ critically conditions the droplet-wall interactions: from purely chemical at low $\Pe$ to hydrochemically coupled at higher $\Pe$~\cite{Lippera2020}. Focusing on the non-axisymmetric motion along a plane wall, Ref.~\cite{Desai2021} recently demonstrated that active drops initially `hovering' near a wall destabilize via an advective instability, just like in unbounded domains. In fact, and maybe somewhat counter-intuitively, self-propulsion is promoted by the presence of a boundary: the critical P\'eclet number above which self-propulsion can develop is reduced monotonically as the drop-wall separation reduces, reaching $\Pe_c \approx 2$ in the limit of vanishing separation, i.e. about half its value as compared to the unbounded swimming case~\cite{Desai2021}. The details of the \emph{long-time} propulsion of active drops along rigid walls, however, remain unclear despite being a quintessential experimental configuration.

\begin{figure}[h]
    \begin{center}
    \subfloat[hovering, non-quiescent state]{\label{schem_base}\includegraphics[width=7cm]{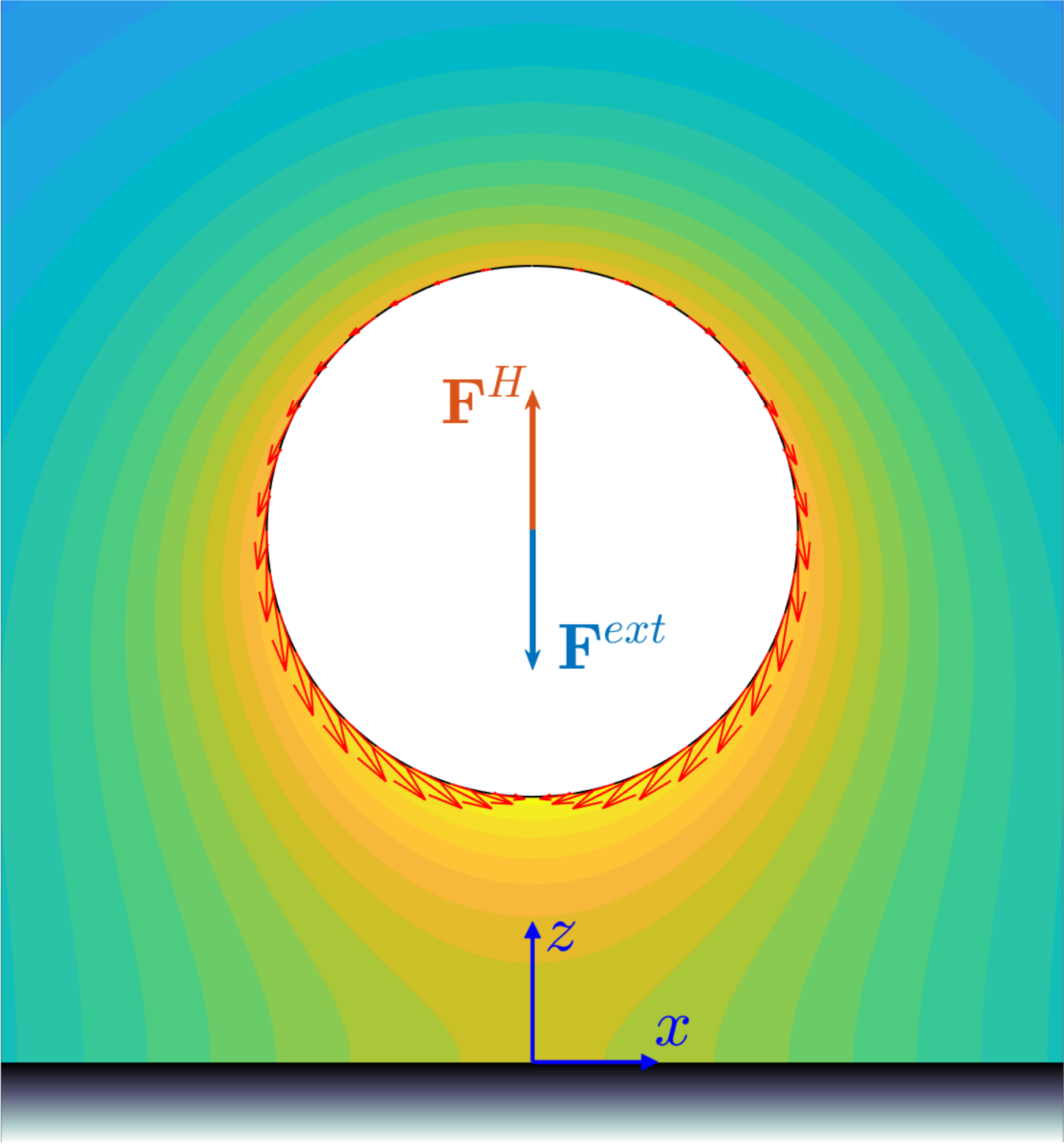}}
    \hspace{5mm}
    \subfloat[steady propulsive state]{\label{schem_prop}\includegraphics[width=7cm]{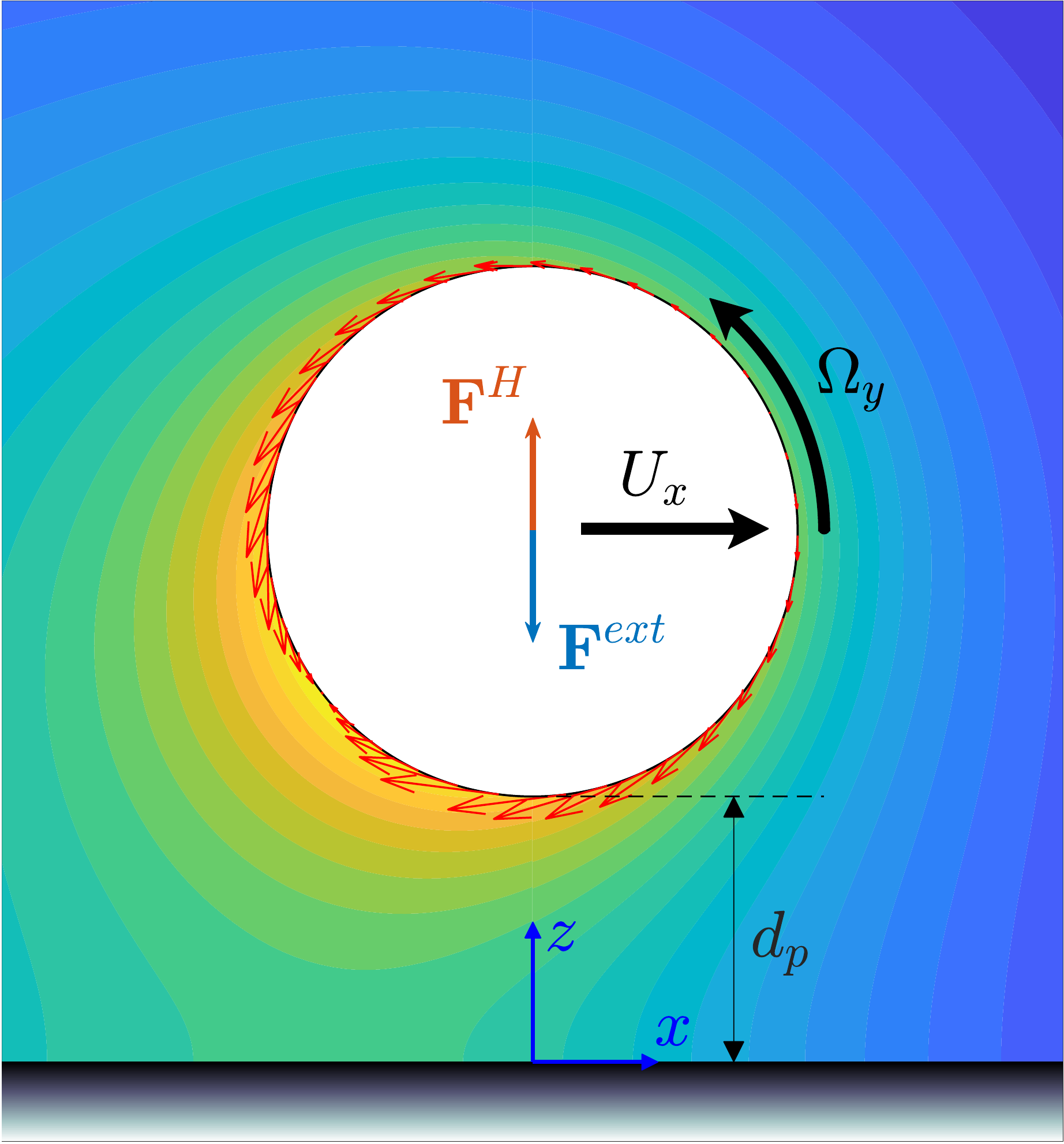}}

    \caption{(a) Hovering and (b) self-propelling states of an isotropic phoretic particle near a rigid wall. The particle-wall separation is set by a balance between the external force $\mathbf{F}^{ext}$, and the solute-accumulation-induced repulsive force $\mathbf{F}^H$.}
    % Different states of an isotropic active particle (activity - $\mathcal{A}$; mobility - $\mathcal{M}$) near a rigid wall.  The particle-wall separation is set by a balance between the external force $\mathbf{F}^{ext}$, and the solute-accumulation-induced repulsive force $\mathbf{F}^H$. Depending on the external force and the P\'eclet number of the system, the particle can exhibit, (a) a stationary, non-quiescent hovering state, or, (b) a steady propulsive state. Red arrows on the particle surface denote local velocity vectors in the particle frame.}
    \label{schematic}
    \end{center}
    \end{figure}

This paper aims to bridge this gap and explores the possibility of long-time self-propulsion of active drops along rigid walls. The propulsion is expected to arise from the advective instability of a stationary, but non-quiescent base state in which the non-neutrally buoyant drop hovers over the wall~\cite{Moerman2017}. The impermeable wall prevents diffusion of the solute emitted by the drop, and generates a vertical concentration contrast across the drop leading to `pumping flow' toward the wall. The resulting hydrodynamic force (away from the wall) on the drop balances the external force (e.g., gravity) acting on it and maintains an axisymmetric equilibrium (see schematic in Fig. \ref{schem_base}). As discussed in Ref.~\cite{Desai2021}, this axisymmetric state may become unstable to longitudinal swimming modes, yielding the self-propulsion we seek to characterize (Fig. \ref{schem_prop}). The propulsion is contingent on the interaction between the drop's activity, its mobility and the \emph{advective} transport of the emitted solute by the mobility-induced flow. One way in which these effects can be realized is through a combination of diffusiophoresis (generation of interfacial `slip' velocity) and Marangoni forcing (generation of interfacial stress), caused by asymmetric distribution of micelles released by solubilizing drops. The relative significance of diffusiophoresis and Marangoni forcing toward driving fluid flow remains obscure, particularly for surfactant-rich drops with immobile interfaces~\cite{Cui2013}. So, in this first exploration we use a purely phoretic approach to droplet propulsion. In addition, we assume the viscosity of the drop to be large as compared to its suspending fluid. In this way, we analyse the motion of an isotropic `active particle' as an approximation of the much more complicated motion of an `active drop'. We generalize the numerical framework of Ref.~\cite{Desai2021}$-$based on bispherical harmonic expansions of hydro-chemical fields$-$beyond the linearised limit to account for the non-linearly coupled fluid and solute transport around an active particle. We then use our numerical method to find steady solutions to the problem of an active particle swimming parallel to a passive rigid wall.

The rest of the paper is organized as follows. Section \ref{mathMod} provides a physical description of the system, followed by a mathematical model of the non-linear chemo-hydrodynamics problem for the active particle. Section \ref{soluMeth} outlines our methodology to obtain the steady swimming solutions to this problem, while the more technical details are included in the Supplementary Material (SM). In Section \ref{results}, we analyse our results and give physical insights into the self-propulsion of the active particle along a rigid wall. \nd{Finally, Section \ref{conclusion} summarizes our study and lists perspectives for future investigation.}

\section{Mathematical model}\label{mathMod}

\subsection{Physical description}\label{mathMod1}

We consider a particle that emits a solute at a constant rate, $\mathcal{A}>0$, due to chemical reactions on its surface. The bulk concentration of this solute is $c^*$ and its molecular diffusivity is $D$. Phoretic effects on the surface of the particle are characterized by a mobility $\mathcal{M}>0$; they cause interfacial flows as a response to gradients in the surface concentration of the emitted solute: $\ub_s^* = \mathcal{M} \nabla_s c^*$. These flows transport the solute via advection and sustain the particle's self-propulsion if the system's P\'eclet number, $\Pe$, is large enough. It must be noted here that while the physical significance of the activity and mobility might change with their respective signs, the spontaneous emergence of self-propulsion remains unchanged provided $\mathcal{AM}>0$ \cite{Michelin2013}.

In the present work, a \emph{fixed} external force (e.g., buoyancy), $-F^{ext} \eb_z$, is acting on the particle, which is balanced by the total hydrodynamic force exerted by the fluid, $F^H_z \eb_z$, resulting from the particle's motion and the wall-induced polarity of solute concentration. So, to quantify the long-time swimming of the active particle, one must simultaneously solve the equations governing fluid flow and solute transport.

\subsection{Governing equations}\label{mathMod2}

We define the characteristic velocity as $V_c = \mathcal{AM}/D$, and non-dimensionalize length, velocity, pressure and time by the scales, $R$, $V_c$, $\eta V_c/R$ and $R/V_c$, respectively, where $\eta$ is the dynamic viscosity of the suspending fluid. The solute concentration is represented in terms of a dimensionless, relative concentration $c = \left( c^* - c^*_{\infty} \right)D/ \left( \mathcal{A} R \right)$. Using these scales, the P\'eclet number is given by $\Pe = V_cR/D = \mathcal{AM}R/D^2$. The dimensionless advection-diffusion equation for the solute being emitted by the active particle is,
\begin{equation}\label{advDiff_general}
\frac{\partial c}{\partial t} + \ub \cdot \nabla c = \frac{\nabla^2 c}{\Pe}.
\end{equation}
The normalized concentration, $c$, satisfies the boundary conditions,
\begin{equation}\label{advDiff_bcs}
    {\left. {\mathbf{n} \cdot \nabla c} \right|}_{\mathscr{W}} = 0, \qquad{\left. {\mathbf{n} \cdot \nabla c} \right|}_{\mathscr{S}} = -1,
\end{equation}
where $\mathbf{n}$ is the outward pointing normal on the surface of the particle, and $\mathscr{W}$ and $\mathscr{S}$ denote the wall and the particle surface, respectively. In addition, the normalized concentration vanishes far away from the particle,
\begin{equation}\label{advDiff_bcs_inf}
    {\left. c \right|}_{r \to \infty} = 0.
\end{equation}

In Eq.~\eqref{advDiff_general}, $\ub$ is the velocity field (in lab frame) of the fluid surrounding the particle, governed by the incompressible Stokes equations:
\begin{equation}\label{continuity}
    \nabla \cdot \ub = 0, \qquad - \nabla p + \nabla^2 \ub = \mathbf{0}.
\end{equation}
with the velocity vanishing at the wall,
\begin{equation}\label{Stokes_bcs_wall}
    {\left. \ub \right|}_{\mathscr{W}} = \mathbf{0},
\end{equation}
and far away from the particle (fluid at rest)
\begin{equation}\label{Stokes_bcs_inf}
    {\left. \ub \right|}_{r \to \infty} = \mathbf{0}.
    \end{equation}
The fluid flow is driven at the particle surface by the local, surface gradient of the solute concentration, and the particle's translation and rotation,
\begin{equation}\label{Stokes_bcs_particle}
    {\left. \ub \right|}_{\mathscr{S}} = \nabla_s c + \mathbf{U} + \mathbf{\Omega} \times \mathbf{x}_s,
\end{equation}
where, $\mathbf{x}_s$ is the position vector from the center of the particle to its surface, and $\mathbf{U}$ and $\mathbf{\Omega}$ are the particle's translational and rotational velocities, respectively. They are obtained by enforcing that the particle must experience zero total force and torque at all times:
\begin{equation}\label{force_int_gen}
    \int_{S} { \mathbf{n} \cdot \boldsymbol\sigma dS } + \mathbf{F}^{ext} = \mathbf{0},\qquad  \int_{S} { \mathbf{x}_s \times \left( \mathbf{n} \cdot \boldsymbol\sigma \right)dS } = \mathbf{0},
\end{equation}
where $\boldsymbol\sigma$ is the stress tensor in the fluid.

\section{Solution methodology}\label{soluMeth}
% The existence of such a solution dictates that Eqns. \eqref{advDiff_general} to \eqref{force_int_gen} must be solved in a reference frame that is \emph{moving} with the particle, since it is only in this frame that the flow and solute concentration fields would appear steady (non-varying with time)
\subsection{The steady state problem}\label{soluMeth1}

We are interested here in steady self-propulsion. In such states, the phoretic particle's velocity is necessarily along the wall as any motion along another direction would result in time-dependent problem as the wall-particle separation is modified. As a result, in the following  we seek solutions of the problem outlined in Section \ref{mathMod2} that are stationary in the co-moving particle frame. The advection-diffusion equation thus reads,
\begin{equation}\label{advDiff_steady}
\left( \ub - \mathbf{U} \right) \cdot \nabla c = \frac{\nabla^2 c}{\Pe},
\end{equation}
where $\left( \ub - \mathbf{U} \right)$ is now the fluid velocity in the body-fixed frame, and $\mathbf{U} = U_x \eb_x$.
% In the steady swimming state, we expect the particle to only translate parallel to the wall, since its vertical motion is prevented by a balance between the external force, $\mathbf{F}^{ext} = -F^{ext} \eb_z$, and the hydrodynamic force in the propulsive state, $\mathbf{F}^H_p = F^H_z \eb_z$. We can thus replace $\mathbf{U} = U_x\eb_x$ in Eqn.~\eqref{advDiff_steady}.
% We use the hydrodynamic boundary conditions, Eqns.~\ref{Stokes_bcs_wall} to \ref{force_int_gen}, to obtain $\ub^i$ as a function of $c^i$.
Eqn.~\eqref{advDiff_steady} is non-linear since $\left( \ub - \mathbf{U} \right)$ is itself a linear function of $c$, determined by the hydrodynamic constraints, Eqns.~\eqref{Stokes_bcs_wall}--\eqref{force_int_gen}. To solve Eqn.~\eqref{advDiff_steady}, the flow and solute concentration fields are expanded in terms of non-axisymmetric, bispherical harmonic eigenfunctions \cite{Lee1980, Mozaffari2016}. Substitution into, and projection of Eqns.~\eqref{advDiff_steady} and \eqref{advDiff_bcs} onto the orthogonal bases of harmonic expansions recasts the system as a set of non-linear coupled differential equations for concentration modes, solved iteratively (see SM for a detailed description). For a given P\'eclet number and propulsion height, $d_p$, the iterations are initialized with $\left( \ub, c \right)$ corresponding to the hovering state of the particle. Additionally, a small non-axisymmetry is imposed by initializing $\mathbf{U} = 10^{-3} \eb_x$.

It is important to note that our numerical framework accepts the propulsion height and P\'eclet number, $\left( d_p, \Pe \right)$, as inputs, and then yields the particle's linear and angular velocities and the vertical hydrodynamic force acting on it, $\left( U_x, \Omega_y, F^H_z \right)$, as outputs. The physical situation, on the other hand, corresponds to the steady swimming state for a fixed external force, $F^{ext}$, and a prescribed P\'eclet number. Thus, while we know the hydrodynamic force on the particle ($F^H_z = F^{ext}$ by vertical force balance), the propulsion height, $d_p$, is not known \emph{a priori} and may depend on the propulsion (or not) of the particle. This necessitates the use of a second iterative scheme where we iterate on $d_p$ until the resulting $F^H_z$ is within an agreeable tolerance of the prescribed $F^{ext}$ (e.g., $\left| (F^H_z - F^{ext})/F^{ext} \right| < 1\%$). In this way, we obtain the propulsion characteristics, $\left( U_x, \Omega_y, d_p \right)$ as a function of $\left( F^{ext}, \Pe \right)$. Note that the convergence of this iteration is based on the observation that the chemically-induced repulsion of the particle decreases (resp. increases) as its separation from the wall increases (resp. decreases)~\citep{Desai2021}. So, for a fixed external force, a particle that is too far from the wall is expected to get attracted to the wall due to the negative force differential ${\left( F^H_z - F^{ext} \right) \eb_z < 0}$; similarly, a particle that is too close to the wall is expected to get pushed away due to the positive force differential $\left( F^H_z - F^{ext} \right) \eb_z > 0$.
% So, for a fixed external force, a particle that is too far from the wall (respectively, too close to the wall) is expected to get attracted to the wall (respectively, repelled from the wall) due to the negative (respectively, positive) force differential $\left( F^{ext} - F^H_z \right) \eb_z$.

\subsection{Validation of numerical method}\label{soluMeth2}

\begin{figure}
    \begin{center}
    \subfloat[]{\label{val1}\includegraphics[width=7.5cm]{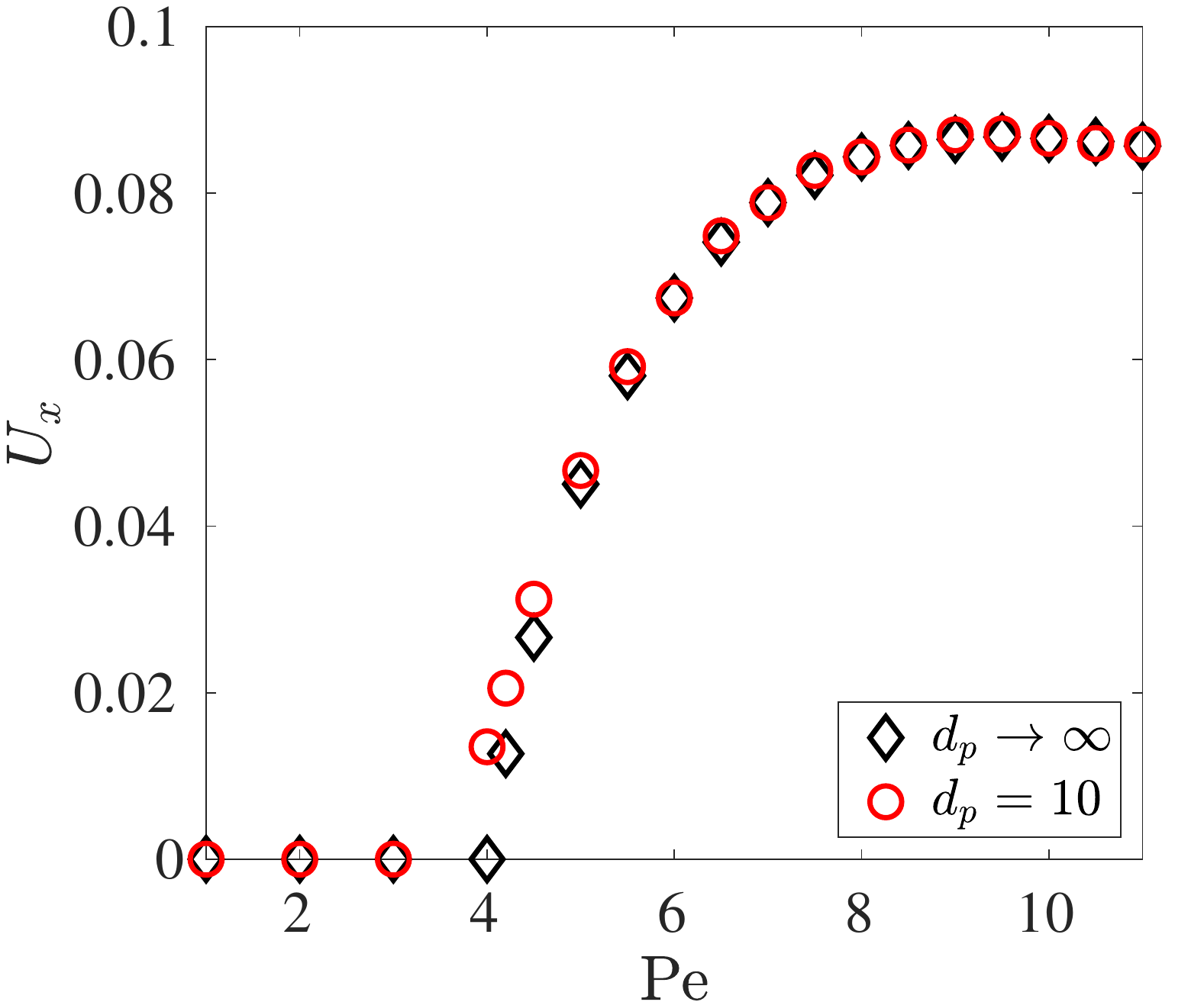}}
    \hspace{5mm}
    \subfloat[]{\label{val2}\includegraphics[width=7.5cm]{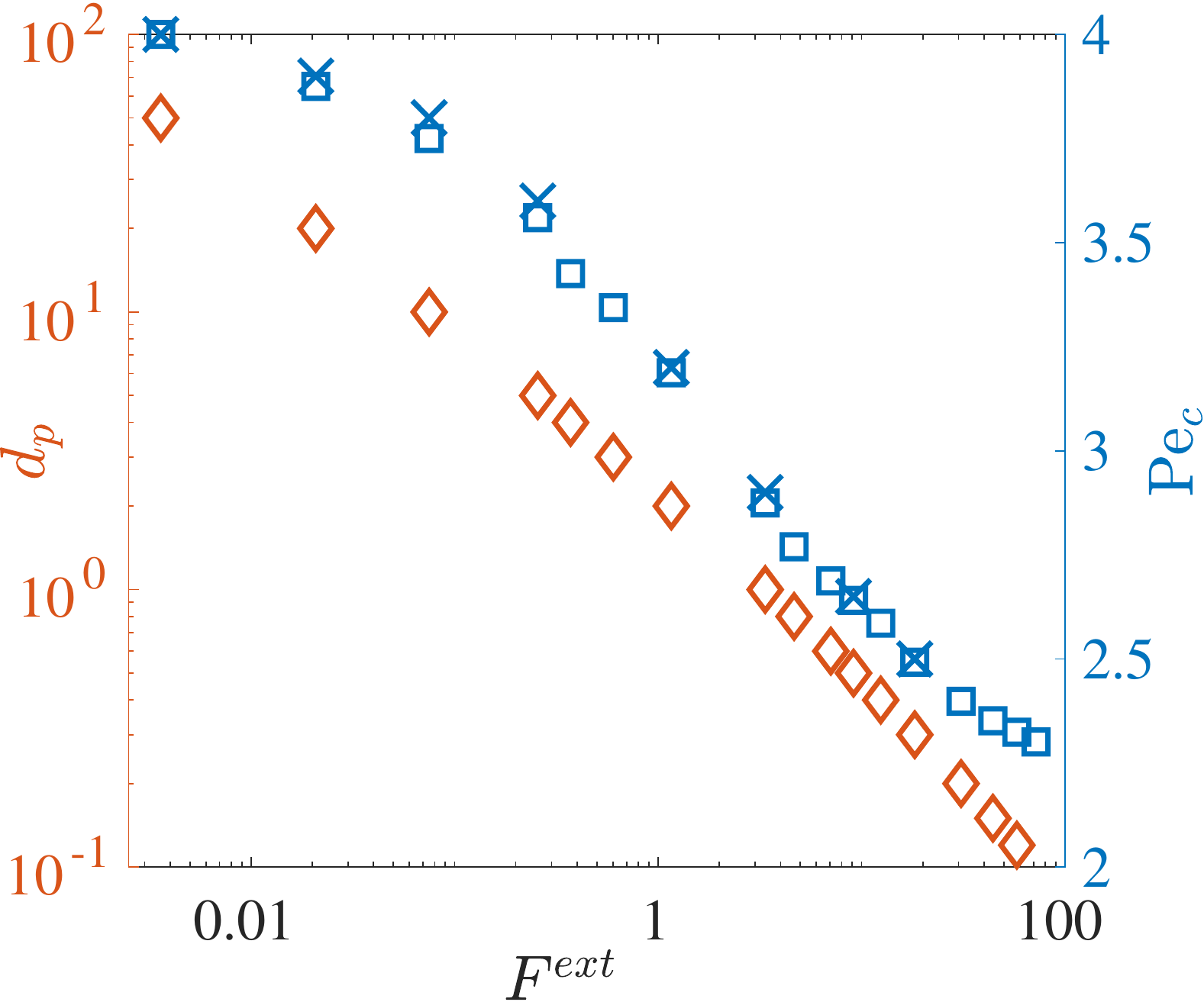}}

    \caption{Validation of the numerical code. (a) Evolution of the swimming velocity $U_x$ as a function of the P\'eclet number $\Pe$ for large particle-to-wall separation and comparison against the results from the unbounded case~\cite{Michelin2013}. (b) Evolution of the critical P\'eclet number for the onset of self-propulsion $\Pe_c$ (right $y$-axis), as a function of the external force $F^{ext}$, acting on the particle. The squares denote results from the linear stability analysis~\cite{Desai2021} and the crosses denote results from the non-linear simulations in the present work. Also shown is the propulsion height $d_p$ (left $y$-axis), corresponding to $\Pe_c$ at which the hovering active particle destabilizes.}
    \label{validations}
    \end{center}
    \end{figure}

For $d_p \gg 1$, the wall is expected to have a negligible influence on the fluid and solute transport, and the swimming velocity, $U_x$, should match that of an isotropic phoretic particle propelling in an unbounded fluid~\cite{Michelin2013}. This comparison is shown in Fig. \ref{val1} and the excellent agreement serves as a first validation of our numerical method.

A second validation is performed by identifying the critical P\'eclet number, $\Pe_c$, above which an active particle hovering at a separation $d_p$ begins to swim spontaneously; we then compare the results to those obtained from linear stability analysis~\cite{Desai2021}. We must however justify this comparison of the results of our non-linear analysis for a \emph{fixed force} $F^{ext}$, against those of the linear stability analysis (for a \emph{fixed propulsion height} $d_p$). Ref.~\cite{Desai2021} showed that, to leading order in the perturbations from an axisymmetric state, the components of the particle's motion normal and parallel to the wall are decoupled. Therefore, for ${\Pe = \Pe_c^+ \equiv \Pe_c + \delta \Pe\;\left(\delta \Pe \ll 1\right)}$, the propulsion of a destabilized particle does not affect, at leading order, the vertical hydrodynamic force it experiences and the particle swims at a separation from the wall which is the same as its initial hovering separation. The constant force and constant separation problems are thus equivalent near $\Pe_c^+$. In this way, we can use the `fixed-separation version' of our numerical framework to identify $\Pe_c$ as a function of $d_p$ and compare against the $\Pe_c(d_p)$ data from Ref.~\cite{Desai2021}. Crucially, since we have \emph{a priori} knowledge of \emph{both} the external force on the particle and its propulsion height, we can transform the $\Pe_c(d_p)$ dependence from our non-linear analysis to a $\Pe_c(F^{ext})$ dependence, where, $F^{ext} = F^H_z(d_p, \Pe_c^+)$. This is plotted in Fig.~\ref{val2} (blue crosses), and compared to the results of the linear stability analysis (blue squares), \nd{where the forces have been made dimensionless by the quantity $\eta V_c R$}. A favorable comparison further confirms the validity and accuracy of our approach.

\section{Characteristics of near-wall propulsion}\label{results}

\begin{figure}
    \begin{center}
    \subfloat[]{\label{Ux_vs_Pe_const_Fz_a}\includegraphics[width=7.5cm]{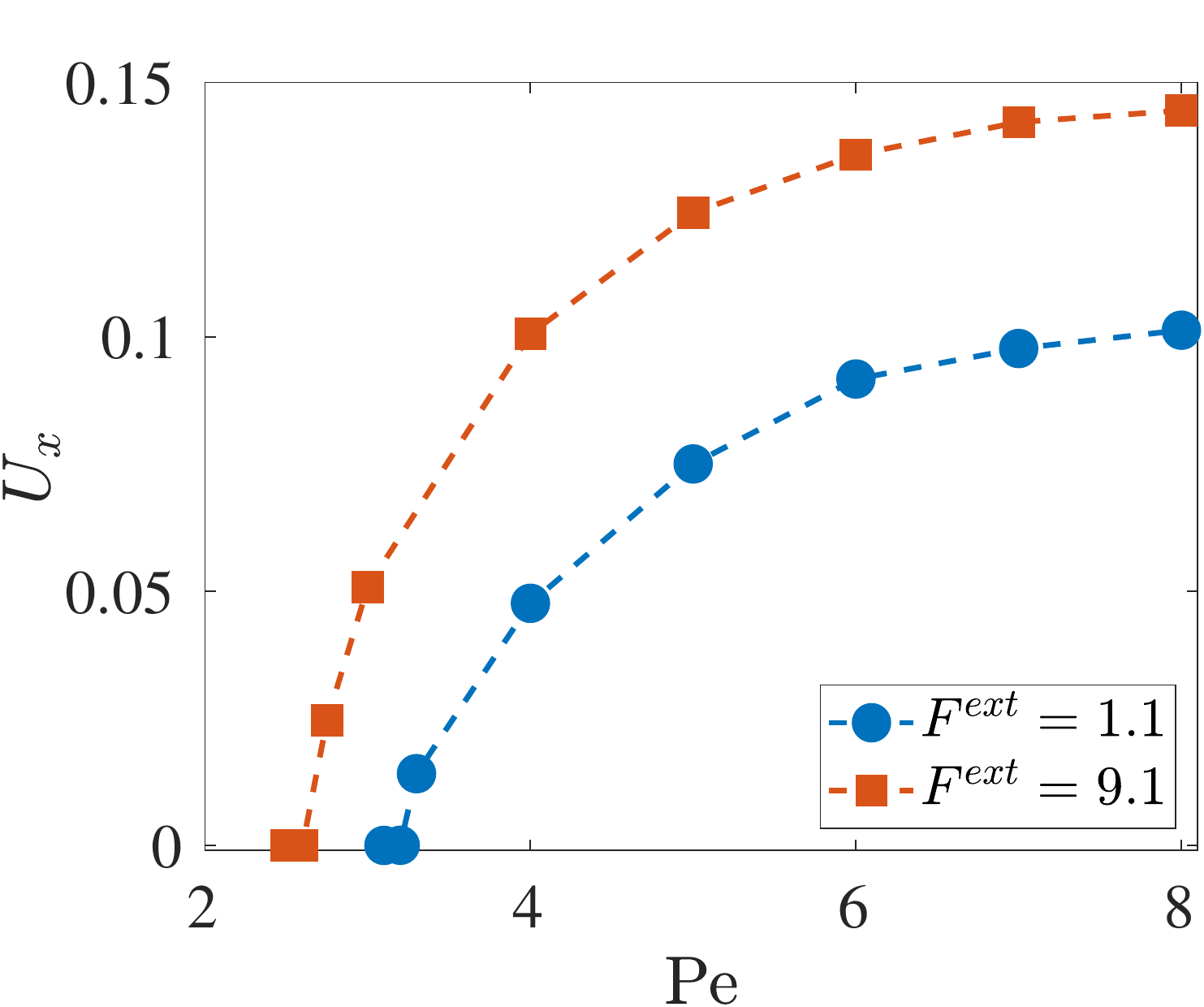}}
    \hspace{5mm}
    \subfloat[]{\label{Ux_vs_Pe_const_Fz_b}\includegraphics[width=7.5cm]{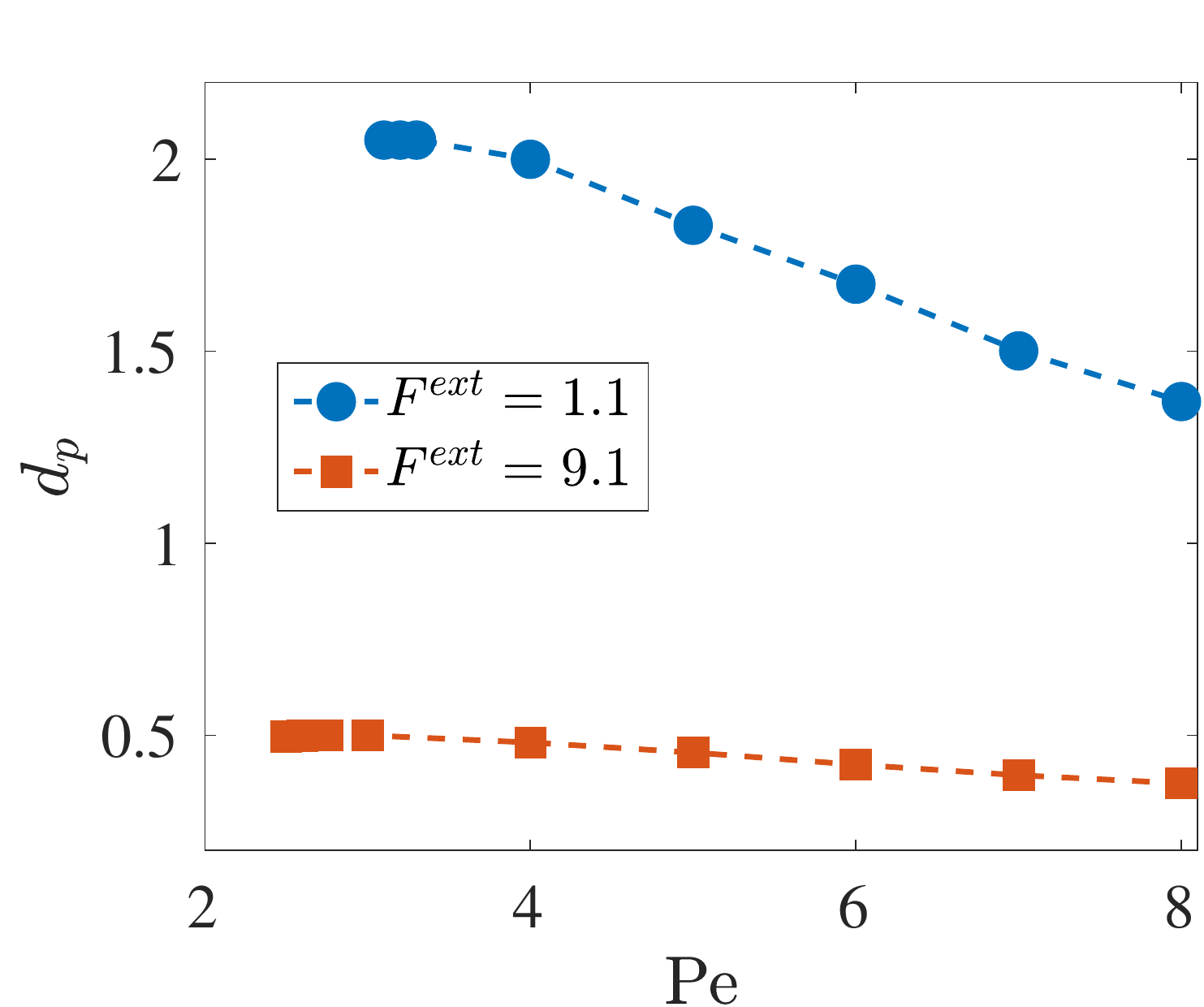}}
    \hspace{5mm}
    \subfloat[]{\label{Ux_vs_Pe_const_Fz_c}\includegraphics[width=7.5cm]{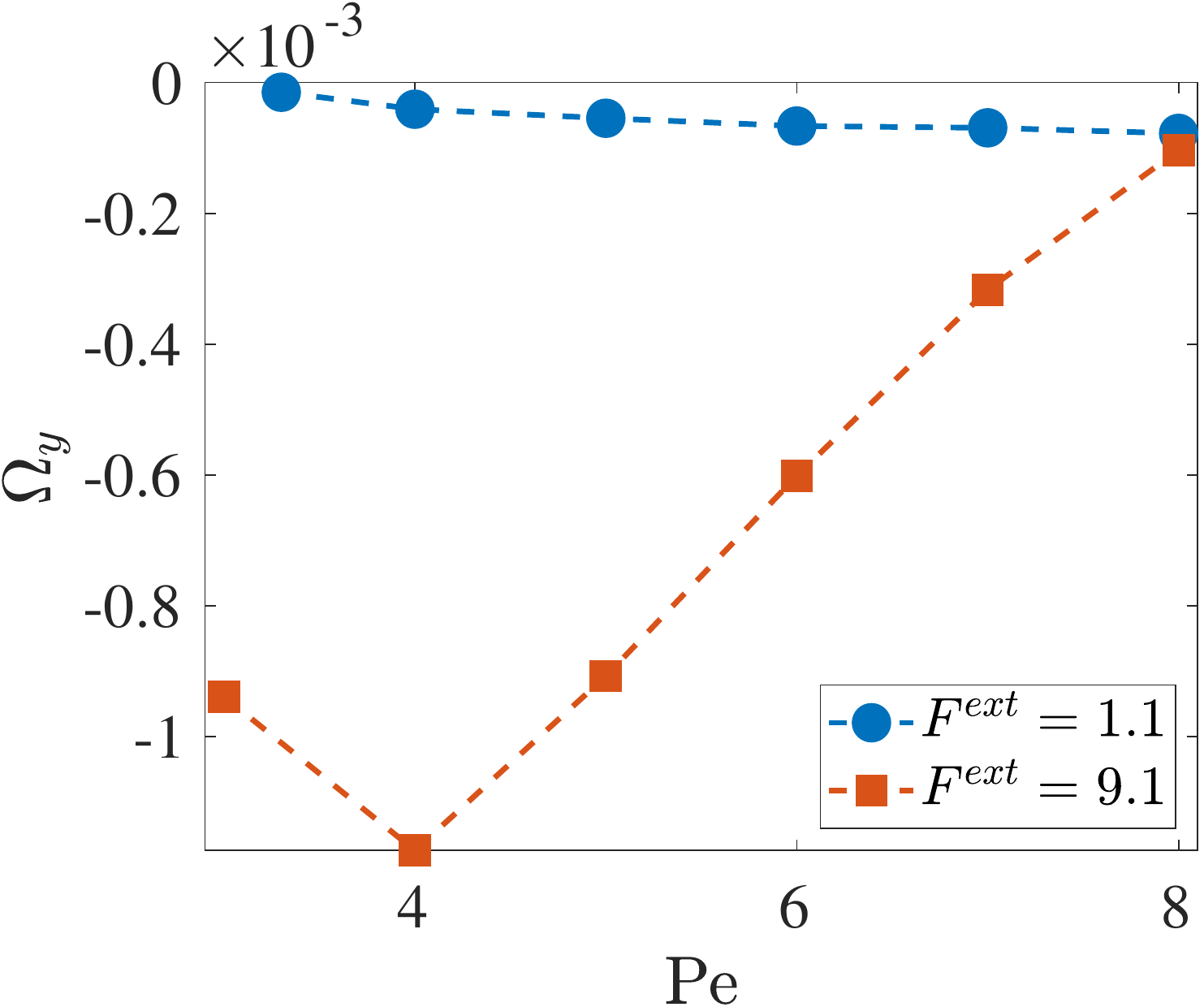}}

    \caption{Evolution with $\Pe$ of (a) the self-propulsion velocity $U_x$, (b) the propulsion height $d_p$, and, (c) the angular velocity $\Omega_y$, for a particle experiencing a fixed external force, $F^{ext}$.}
    \label{Ux_vs_Pe_const_Fz}
    \end{center}
    \end{figure}

We now analyse the evolution of the propulsion characteristics, $\left( U_x, d_p, \Omega_y \right)$, with $\Pe$ and increasing values of a fixed external force (Figure~\ref{Ux_vs_Pe_const_Fz}).

\subsection{Steady state translation}\label{results1}

\begin{figure}
    \begin{center}
    \subfloat[]{\label{de_vs_dp_b1}\includegraphics[width=7cm]{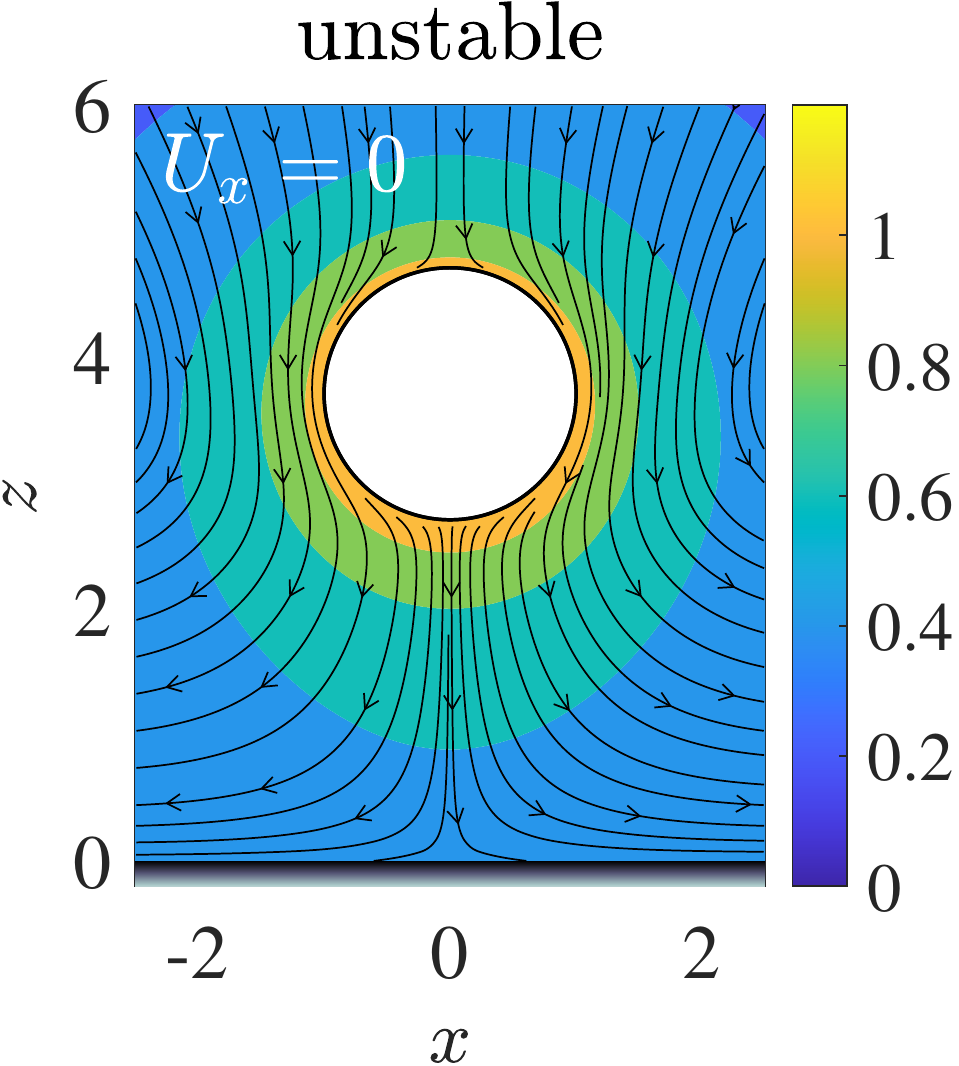}}
    \hspace{5mm}
    \subfloat[]{\label{de_vs_dp_b2}\includegraphics[width=7cm]{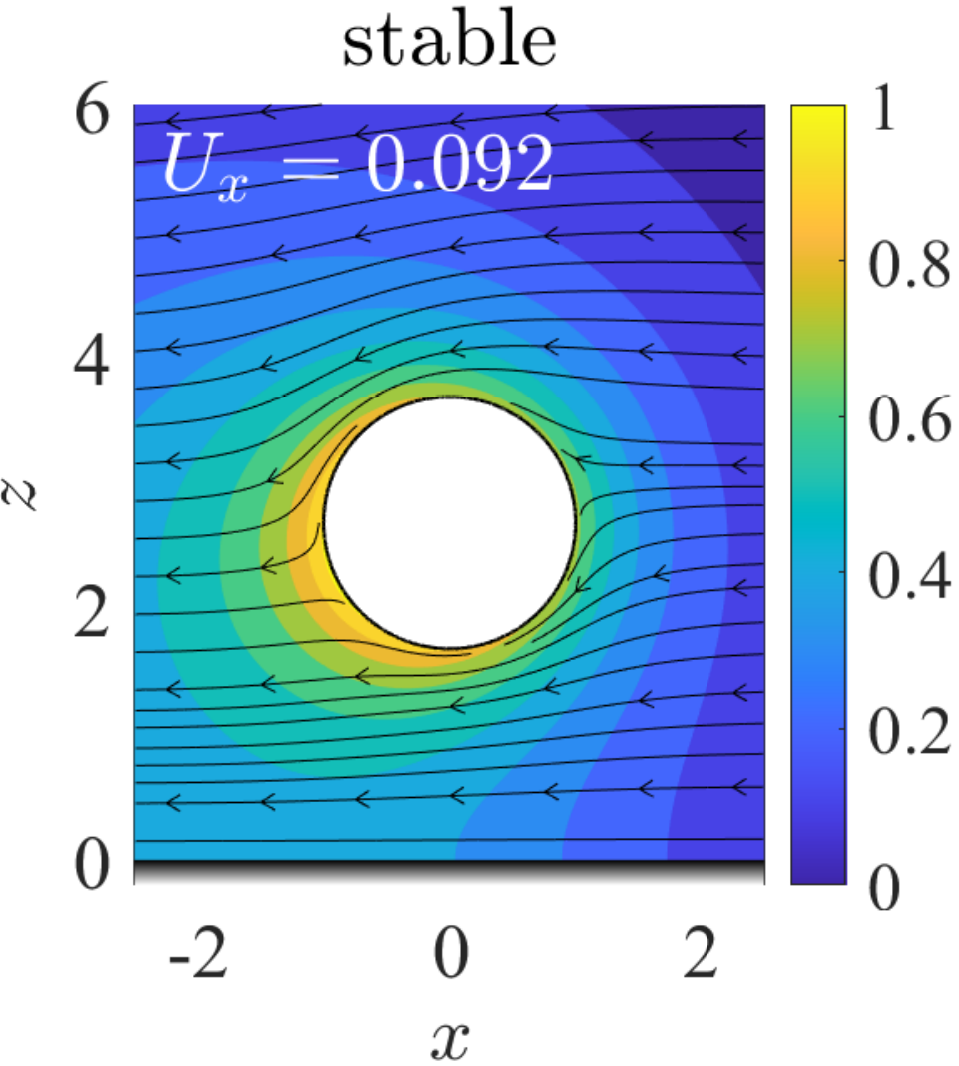}}

    \caption{(a) Unstable hovering state and (b) stable propulsive state of a phoretic particle experiencing a constant force, $F^{ext}=1.1$ for $\Pe=6$. The solute concentration (color) and streamlines (black lines) are shown in each case.}
    \label{de_vs_dp_b}
    \end{center}
    \end{figure}

The translational velocity, $U_x$, first increases with respect to $\Pe$ but seems to saturate for larger $\Pe$ as for the unbounded case~\citep{Michelin2013} (Figure~\ref{Ux_vs_Pe_const_Fz_a}). Interestingly,  the propulsion height, $d_p$, \emph{reduces} with $\Pe$. As the axisymmetric base state destabilizes, the particle begins to move along $\eb_x$ and the solute in the particle-wall gap gets advected downstream (i.e., along $-\eb_x$). The vertical polarity is reduced (and so is the effective repulsive interaction with the wall resulting from solute accumulation): the particle thus `descends' under the influence of $F^{ext}$ until the vertical force balance is restored and a steady propulsion regime is achieved (see Fig. \ref{de_vs_dp_b}). \nd{The characteristic (dimensional) time scale for such descent is $t_d \sim R/V_c$ with $V_c=\mathcal{AM}/D$ (i.e. $t_d=O(1)$ in non-dimensional form here), resulting from the balance between  the external force and the chemically-induced wall-repulsion, and also indicative of the time taken by the longitudinal flow to `sweep' the solute from underneath the particle to behind the particle.} Stronger advection is able to drive away more solute from the particle-to-wall gap and thus causes further lowering of the particle, i.e., $d_p$ reduces for increasing $\Pe$ (as seen clearly for $F^{ext} = 1.1$ in Fig.~\ref{Ux_vs_Pe_const_Fz_b}).

To explain the evolution of the swimming speed with  $\Pe$, $U_x$ can be computed as a weighted surface average of the forcing concentration gradient using the reciprocal theorem~\citep{Stone1996},
\begin{equation}\label{Ux_reciproc}
    U_x = \int_{S} { \mathbf{n} \cdot \boldsymbol {\hat\sigma}_T \cdot \nabla_s c\;dS },
\end{equation}
where $\mathbf{n} \cdot \boldsymbol{\hat\sigma}_{T}$, the influence function, is the hydrodynamic traction on the particle in a carefully-chosen auxiliary Stokes flow problem. The influence function is a measure of the effectiveness of the surface slip toward self-propulsion, and as such depends upon the location on the particle surface where the slip acts and its hydrodynamic environment. To obtain $U_x$, the auxiliary Stokes problem considered to obtain $\mathbf{n} \cdot \boldsymbol{\hat\sigma}_T$ is the translational and rotational motion of a rigid torque-free particle under the influence of a unit external force, $\mathbf{F}^{ext}_T = \eb_x$. 

\begin{figure}
    \begin{center}
    \includegraphics[width = \linewidth]{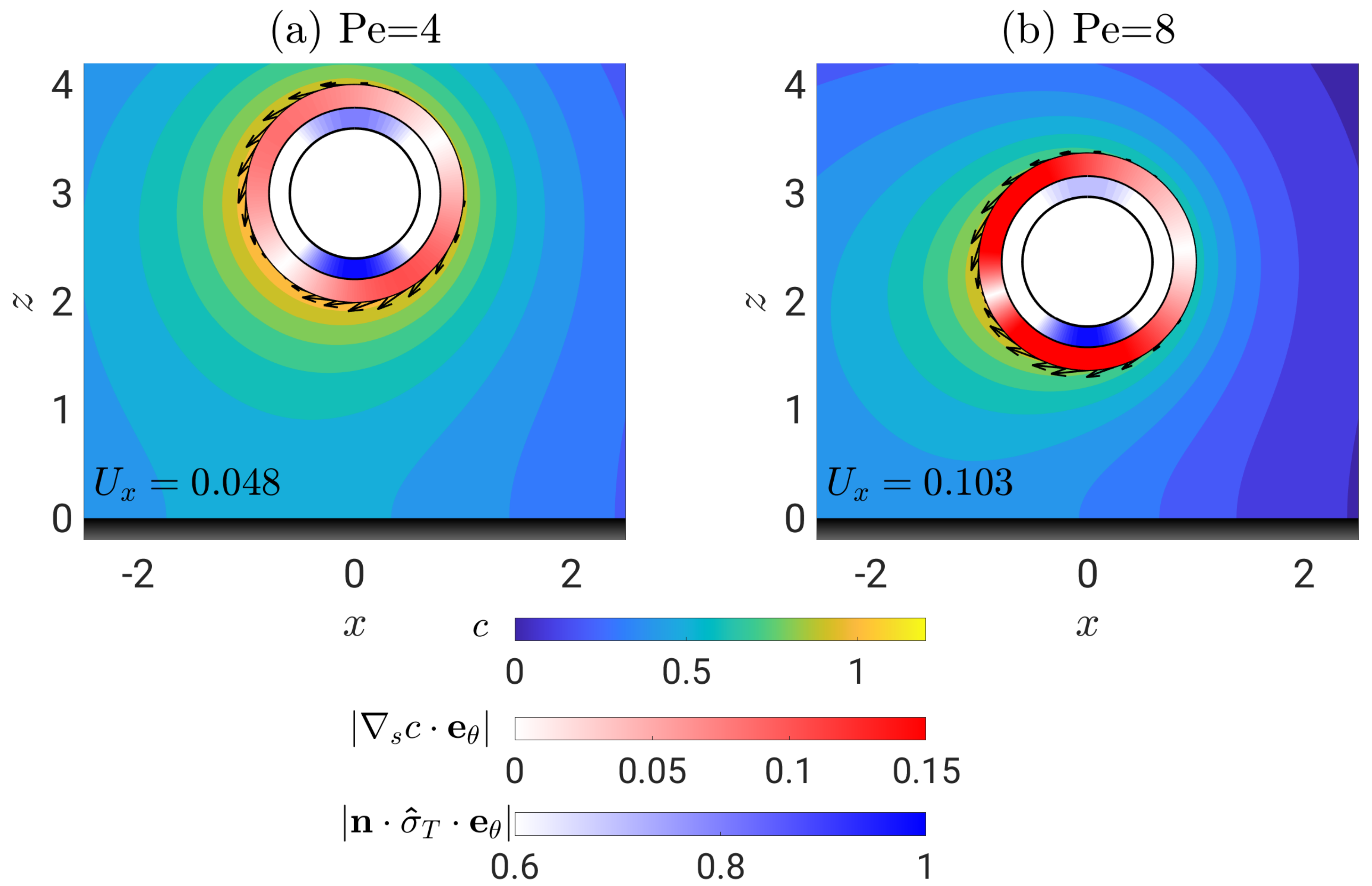}

    \caption{Azimuthal surface gradient in solute concentration, $\left| \nabla_s c \cdot \eb_{\theta} \right|$ (white-red outer polar counter), and its normalized influence on the translation velocity of the particle, $\left| \mathbf{n} \cdot \boldsymbol{\hat\sigma}_T \cdot \eb_{\theta} \right|$ (white-blue inner polar counter),  within the $y=0$ plane of symmetry. \nd{The particle surface can be identified by the vectors denoting the slip velocity.} The results are shown here for $F^{ext} = 1.1$ and, (a) $\Pe = 4$, (b) $\Pe = 8$.}
    \label{influence_tracn_trans}
    \end{center}
    \end{figure}

Fig. \ref{Ux_vs_Pe_const_Fz_b} showed that an increase in $\Pe$ causes a progressive reduction in the propulsion height, $d_p$. As the particle nears the wall for larger $\Pe$, the surface concentration gradient $\nabla_s c$, and its influence $\mathbf{n} \cdot \boldsymbol{\hat\sigma}_T$, \emph{both} get localized around the gap between the particle and the wall (see Fig. \ref{influence_tracn_trans}). This efficient `placement' of the surface slip around the particle is the reason for stronger propulsion speeds at the larger P\'eclet numbers considered here. In fact, the combined localization of chemical gradients and fluid traction also leads to more rapid destabilization of the hovering states that are closer to the wall, as was shown in the linear stability analysis in Ref.~\cite{Desai2021}. We have thus confirmed that earlier destabilization of the base state and faster swimming speeds in the steady state both originate in the same physico-chemical phenomena.

\begin{figure}
\begin{center}

\subfloat[$F^{ext} = 1.1$]{\label{prop_Fz_low}\includegraphics[width=7cm]{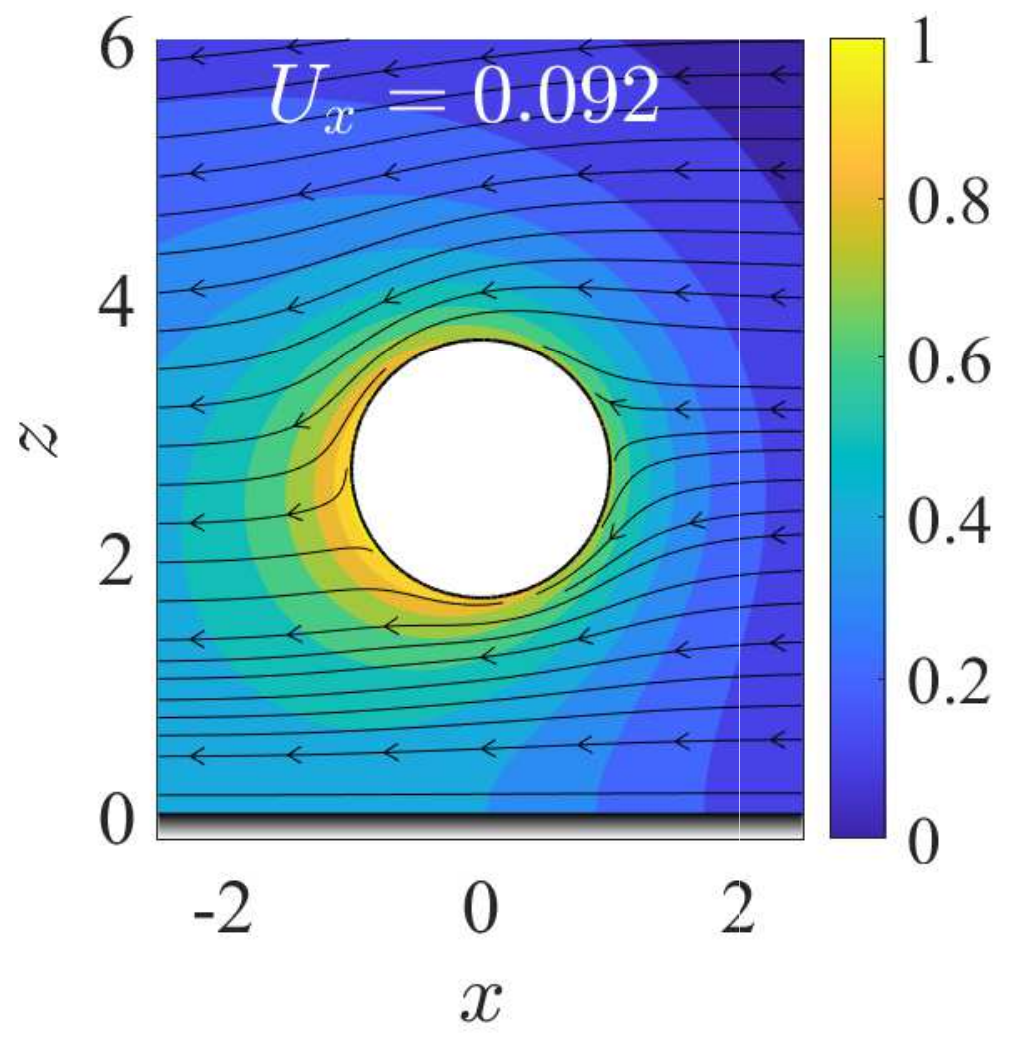}}
\hspace{5mm}
\subfloat[$F^{ext} = 9.1$]{\label{prop_Fz_high}\includegraphics[width=7cm]{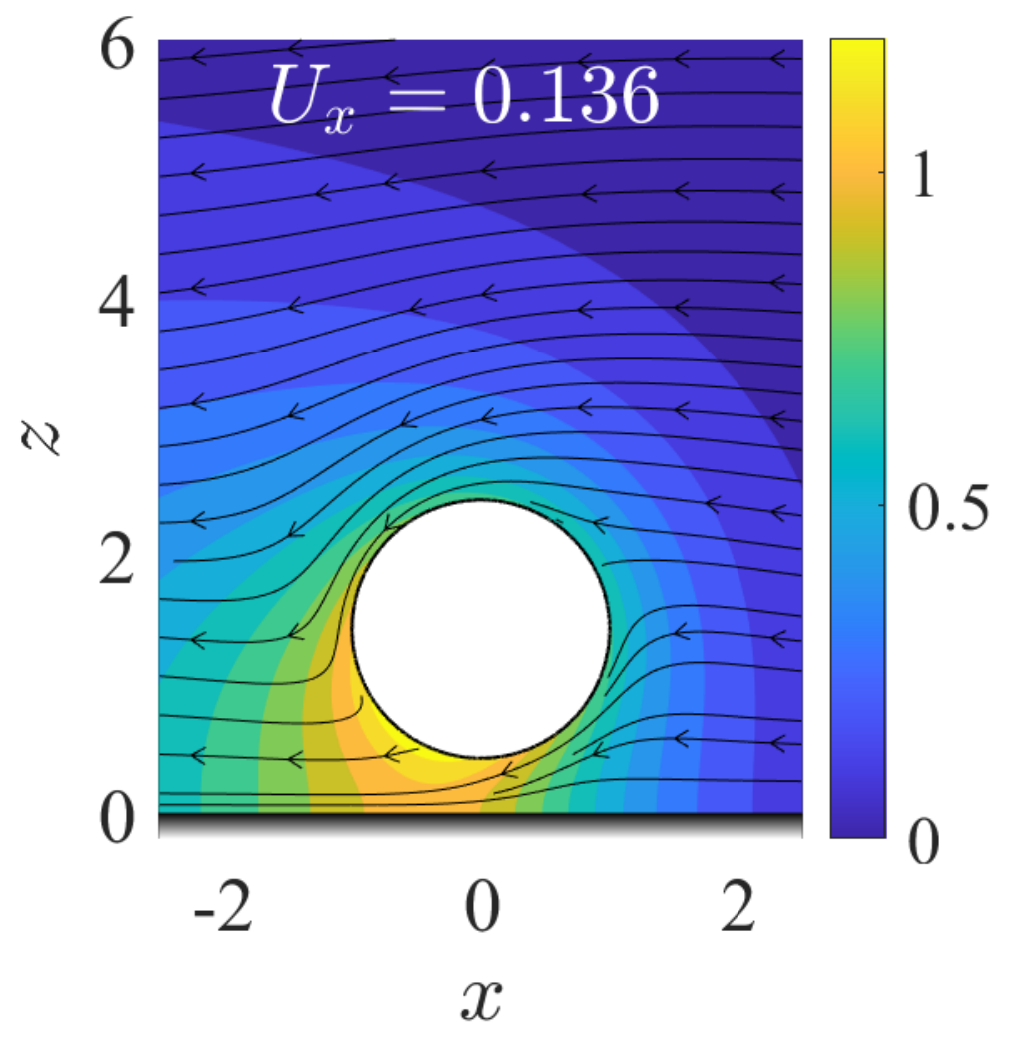}}

\caption{Propulsive states corresponding to two different values of $F^{ext}$, and $\Pe = 6$. Note that higher external force results in stronger $z$-polarity of solute concentration.}

\label{prop_Fz_high_vs_low}

\end{center}
\end{figure}

A higher external force causes the particle to swim closer to the wall in order to generate sufficiently strong vertical polarity in the solute concentration and maintain the force balance (see Fig. \ref{prop_Fz_high_vs_low}). Hence, the surface-slip-driven motion is more efficient and the particle speed increases (Fig. \ref{Ux_vs_Pe_const_Fz_a}).
At higher $F^{ext}$ however, the reduction in $d_p$ with increasing $\Pe$ is much more gradual (Fig. \ref{Ux_vs_Pe_const_Fz_b}). This is due to a competition between solute advection along two orthogonal directions: (i) longitudinally along $-\eb_x$ because of self-propulsion, versus, (ii) normally along $-\eb_z$ because of the vertical flow caused by solute accumulation in the particle-wall gap. For low $F^{ext}$, longitudinal advection of the solute is much more dominant and we see a steeper reduction $d_p$; whereas for high $F^{ext}$, the wall-normal solute advection exerts a stronger influence in opposing the downstream solute advection, and so an increase in the P\'eclet number yields only a modest reduction in the propulsion height.

\subsection{Rotation of the particle}\label{results2}

\begin{figure}
    \begin{center}
    \includegraphics[width = \linewidth]{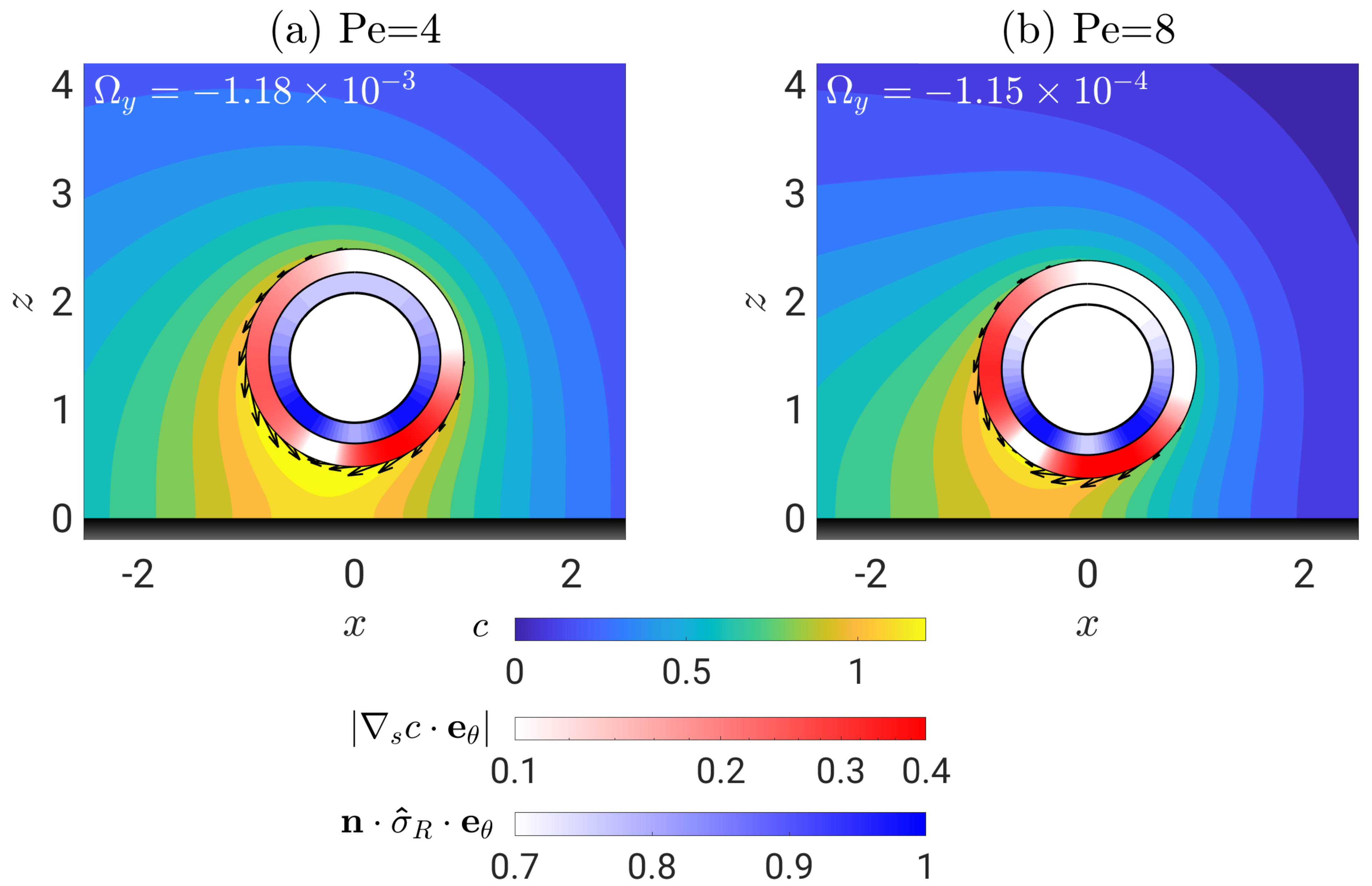}

    \caption{Azimuthal surface gradient in solute concentration, $\left| \nabla_s c \cdot \eb_{\theta} \right|$ (white-red outer polar counter), and its normalized influence on the angular velocity of the particle, $\mathbf{n} \cdot \boldsymbol{\hat\sigma}_R \cdot \eb_{\theta}$ (white-blue inner polar counter) within the $y=0$ plane of symmetry. \nd{The particle surface can be identified by the vectors denoting the slip velocity.} The results are shown here for $F^{ext} = 9.1$ and, (a) $\Pe = 4$, (b) $\Pe = 8$.}
    \label{influence_tracn_rot}
    \end{center}
    \end{figure}

An isotropic phoretic particle translating in an unbounded domain does not rotate since the flow is symmetric around the translational axis of the particle. However, in the presence of a wall, there is an asymmetry of the tangential velocity on the particle surface, across the horizontal plane, which yields an angular velocity along $\eb_y$ of the torque-free particle. The particle's rotation though, is negligibly small for the separations considered here, especially for small values of $F^{ext}$, where  $\left| \Omega_y \right| \sim O(10^{-5})$ (Fig. \ref{Ux_vs_Pe_const_Fz_c}). Interestingly, in all cases, the particle does not ``roll" above the wall as it swims forward, i.e. $\Omega_y < 0$. This is different from the well-known ``rolling" of a rigid sphere moving parallel to the wall under the action of a constant, horizontal external force \cite{Lee1980}. This negative angular velocity can again be explained using the reciprocal theorem to express the particle's angular velocity as
\begin{equation}\label{Omy_reciproc}
    \Omega_y = \int_{S} { \mathbf{n} \cdot \boldsymbol {\hat\sigma}_R \cdot \nabla_s c\;dS },
\end{equation}
where $\mathbf{n} \cdot \boldsymbol{\hat\sigma}_R$ is now the surface traction for the motion (i.e., rotation and translation) of a force-free rigid particle under the influence of a unit external torque, $\mathbf{L}^{ext}_R = \eb_y$. The concentration gradients in the azimuthal direction and their influence on rotating the particle are shown in the polar contour plots in Fig. \ref{influence_tracn_rot}. \nd{In the notation of Fig. \ref{influence_tracn_rot}, $\eb_{\theta}$ is the tangent unit vector on the drop's surface, with $\theta$ measured counterclockwise from $\eb_x$; the fluid traction associated with a rotation along $\eb_y$, $\textbf{n} \cdot \boldsymbol{\hat\sigma}_R \cdot \eb_{\theta}$, is thus always positive. Eqn.~\eqref{Omy_reciproc} then tells us that any clockwise (resp. counterclockwise) concentration gradients on the surface of the particle will contribute negatively (resp. positively) toward the angular velocity, $\Omega_y$. For $\Pe = 4$, the largest surface gradients reside at the front of the particle, near its bottom pole and are oriented clockwise (Fig. \ref{influence_tracn_rot}a; outer ring materializing the particle surface). They also overlap appreciably with the region where the influence function, $\textbf{n} \cdot \boldsymbol{\hat\sigma}_R \cdot \eb_{\theta}$, is maximum (Fig. \ref{influence_tracn_rot}a; inner ring). These combined effects tend to rotate the particle such that $\Omega_y < 0$. For $\Pe=8$, the strongest surface gradients and the region of their maximum influence are not as much aligned as for $\Pe = 4$. Overall, this leads to a reduction in the magnitude of the angular velocity, $\left| \Omega_y \right|$ as observed in Fig.~\ref{Ux_vs_Pe_const_Fz_c}.}

% , but the influence of the slip in front of the particle on its rotation is reduced. There also develops a `counterclockwise' slip (when seen along $\eb_y$) behind the particle, which negates some of the `clockwise' slip and causes a reduction in the magnitude of the angular velocity, $\left| \Omega_y \right|$ as observed on Fig.~\ref{Ux_vs_Pe_const_Fz_c}.
    
\section{Conclusion and Perspectives}\label{conclusion}

The presence of boundaries is a ubiquitous feature in experiments on chemically-active swimming droplets, but is too often ignored in theoretical analyses. This gap was addressed here by analysing the steady, wall-parallel motion of an isotropic phoretic particle confined to a rigid wall by an external force, $F^{ext}$, by solving the complete (non-linear) hydro-chemical problem. Primarily, we demonstrated that the flow around the propelling particle weakens the solute-induced repulsion it experiences from the wall, and necessitates reduced particle-wall separations in order to maintain the vertical force balance. This creates a very efficient surface distribution of the phoretic slip and thus augments the swimming speed of the particle.

An important simplification in our analysis is the assumption of a purely phoretic response to chemical inhomogeneities, while in reality, active drops employ some combination of phoretic and Marangoni responses. The present framework could easily be extended to treat this more generic situation in non-axisymmetric bispherical coordinates. Including the Marangoni boundary condition may change these conclusions, especially the critical P\'eclet number for spontaneous propulsion and the dependence of the propulsion velocity on drop-to-wall separation; yet, a few preliminary remarks can be made. In our study, the localization of the influence of the surface slip near the bottom pole of the particle is a consequence of a `prescribed velocity'/no-slip boundary condition at both the wall and at the particle surface. If the latter is changed into a stress balance condition (as would be the case for a clean drop), then the influence of surface slip is likely not focused at the bottom pole and is more spread out along the surface of the drop. As a result, enhanced wall proximity would not necessarily mean stronger propulsion. The same reasoning also holds for the motion of a phoretic particle near a fluid-fluid interface. This suggests interesting changes can be brought about by relaxing the rigidity constraint of the drop and/or the wall, and these should be analyzed systematically in a future work. However, for drops that are more viscous than their surrounding fluid, phoresis would be the only way to effect fluid flow as a response to chemical gradients and thus our results would be applicable in that case.

We focused here on the \emph{steady} swimming of an active particle, which constrained the particle to move parallel to the wall. Our numerical analysis can be combined with that of Ref.~\cite{Lippera2020}, which utilized deforming bispherical grids, to investigate the interaction between wall-normal and wall-parallel motion of the active particle. We also did not comment on the stability of the obtained swimming states but note that secondary instabilities may arise due to perturbations along the direction orthogonal to both the velocity of the particle and the external force acting on it (i.e., along $\eb_y$ in our coordinate system).

The present numerical method is based on a bispherical harmonic expansion of the flow and concentration fields, and projection of the advection term, $\ub \cdot \nabla c$, onto the orthogonal basis functions of the bispherical expansions. \nd{This results in a non-linear coupling of the polar and azimuthal components of the solute concentration field. In the steady swimming regime, the coupling is strengthened for: (i) large values of $\Pe$, and, (ii) small values of the propulsion separation, $d_p$. Note that the latter condition itself is not independent of the largeness of $\Pe$ and/or $F^{ext}$. An enhanced coupling between the concentration modes thus demands a concomitant increase in the number of terms that need to be retained in the bispherical harmonic expansion summations. As a result, the analysis of larger $F^\text{ext}$ or larger $\Pe$ become rapidly prohibitive; exploring these regimes  with the current numerical framework is therefore not practical, and a different numerical method should be used (e.g., see Ref.~\cite{Picella2022})} to unveil possible near-wall analogues of droplet behaviour observed in unbounded flows, like pausing and reversal~\cite{Morozov2019} or chaotic motion~\cite{Hu2022}. Nevertheless, by modeling the full, non-linear chemo-hydrodynamics of an isotropic active particle, we have illuminated important features of near-wall swimming of isotropic colloids that would otherwise remain elusive in any linearised analysis. We have thus laid the groundwork for a host of prospective theoretical studies on more general features of confined, self-propelling active drops.

% The resulting coupling of the polar and azimuthal components of the concentration field, which is particularly strong for smaller separations, places a large computational demand on the number of modes required to obtain converged solutions.

\begin{acknowledgments}
This work was supported by the European Research Council (ERC) under the European Union's Horizon 2020 research and innovation program (Grant Agreement No. 714027 to S.M.).
\end{acknowledgments}

% \section*{Appendix/SI}

%     \begin{figure}
%     \begin{center}
%     \includegraphics[width=9cm]{Ux_vs_Pe_validation.pdf}
%     \caption{Comparison between the spontaneous propulsion velocity of an isotropic active particle swimming in an unbounded medium ($d_p \to \infty$), with that of an isotropic active particle swimming at a fixed, large separation, $d_p = 10$, from a rigid wall.}
%     \label{Ux_vs_Pe_val}
%     \end{center}
%     \end{figure}

%\bibliographystyle{unsrt}
%\bibliography{apssamp.bib}% Produces the bibliography via BibTeX.
\providecommand{\noopsort}[1]{}\providecommand{\singleletter}[1]{#1}%

\end{document}